\documentclass[12pt]{iopart}
\usepackage{latexsym}

\def\cR{{\cal R}}

\def\Tr{{\rm Tr}}
\def\d={\buildrel \rm def \over =}

\def\ket#1{\mid~\!\!\!{#1}~\!\!\rangle}
\def\bra#1{\langle~\!\!{#1}~\!\!\!\mid}

\begin{document}

\title[\it Twins in entangled mixed states]
{\bf On twin observables in entangled
mixed states}

\author{F Herbut\footnote[1]{E-mail:
fedorh@infosky.net}}
\address{Faculty of Physics, University of
Belgrade, POB 368, Belgrade 11001,
Yugoslavia and Serbian Academy of
Sciences and Arts, Knez Mihajlova 35,
11000 Belgrade}

\date{\today}

\begin{abstract}
It is pointed out that every
mixed-state statistical operator is, up
to a normalization constant, a super
state vector in the Hilbert space of
linear Hilbert-Schmidt operators acting
in the state space of the quantum
system. Hence, the well understood
Schmidt canonical expansion of ordinary
state vectors can be carried over to
mixed states. In particular, it can be
utilized for evaluating all the twins,
i. e., the opposite-subsystem
observables the measurement of one of
which is, on account of entanglement,
{\it ipso facto} also a measurement of
the other. This is illustrated in full
detail in the case of the Horodecki two
spin-one-half-particle states with
maximally disordered subsystems.
\end{abstract}

\submitted \maketitle \normalsize

\section{Introduction}

\rm Recently twin observables, which
make possible distant measurement, i.
e., distant orthogonal state
decomposition on account of
entanglement, have been extended from
the pure state case \cite{FV76} to the
mixed states \cite{HD00}. The
importance of twins for entanglement
studies, in particular for quantum
communication and quantum information
theory, was pointed out.

Statistical operators (physically:
quantum states) are elements of the
Hilbert-Schmidt space of operators or
supervectors, in which the scalar
product is defined by $(A,B)\equiv  \Tr
A^{\dagger}B$. Hence, in many cases the
very well understood twin theory in
pure states can be formally carried
over to the mixed state case. This is
illustrated by an example.

We make use of {\it the Schmidt
canonical expansion} of an arbitrary
state vector $\ket{\Phi}_{12}$ of a
composite system \cite{FV76}. It is
expressed in terms of its {\it reduced
statistical operators} (subsystem
states) $\rho_1$ $\Big(\equiv
\Tr_2\ket{\Phi}_{12}\bra{\Phi}_{12}\Big)$
and $\rho_2$ (defined symmetrically),
and their spectral forms
$$\rho_1=\sum_ir_i\ket{i}_1\bra{i}_1,
\quad
\rho_2=\sum_ir_i\ket{i}_2\bra{i}_2,
\quad \forall i:\enskip r_i>0.
\eqno{(1a,b)}$$
(Note that the positive
spectra are always equal for pure
states.) Further, the mentioned
expansion utilizes the (antiunitary )
{\it correlation operator} $U_a$, which
maps the range $\cR(\rho_1)$ onto the
range $\cR(\rho_2)$. (Note that they
are always equally dimensional for pure
states.) The correlation operator is
determined by $\ket{\Phi}_{12}$, and,
in conjunction with $\rho_1$, it
determines back $\ket{\Phi}_{12}$. The
{\it Schmidt canonical expansion}
reads:
$$\ket{\Phi}_{12}=\sum_ir_i^{1/2}
\ket{i}_1\otimes
\Big(U_a\ket{i}_1\Big)_2. \eqno{(2)}$$
The normalized vectors $\ket{i}_2$ in
(1b) may (and need not) be equal to
$U_a\ket{i}_1$.

For further use it is important to note
that all the mentioned canonical
entities of $\ket{\Phi}_{12}$ are
simply {\it read off} relation (2),
because the Schmidt canonical expansion
is any {\it biorthogonal expansion in
state vectors with positive expansion
coefficients}.

Hermitian twin operators (twin
observables) or simply {\it twins}
$A_1$ and $A_2$ are opposite-subsystem
observables characterized by the
algebraic relation
 $$ A_1
\ket{\Phi}_{12}=A_2 \ket{\Phi}_{12},
\eqno{(3)}$$ where $A_1$ stands for
$A_1\otimes I_2$, and $I_2$ stands for
the second subsystem identity operator,
etc. It was shown that a Hermitian
operator $A_1$ {\it is a twin if and
only if it commutes with} $\rho_1$. The
corresponding twin $A_2$ then satisfies
$$ A_2=U_aA_1U_a^{-1}Q_2+
A_2Q_2^{\perp}, \eqno{(4)}$$
where
$Q_2$ is the range projector of
$\rho_2$, and $Q_2^{\perp}$, its
orthocomplementary projector, projects
onto the null space of $\rho_2$.

One should note that, on account of
symmetry, also $[A_2,\rho_2]=0$, hence
both the range and the null space of
$\rho_2$ are invariant for $A_2$.
Further, the second term on the RHS of
(4) (the component in the null space)
is completely arbitrary and immaterial
for the twin property (3), because it
acts as zero on $\ket{\Phi}_{12}$.
(Naturally, the symmetric claim holds
true for $A_1$ and $\rho_1$.)

\section{MDS states}

 Now we turn to the example that is,
 for illustrative purposes, investigated
in this study, i. e., to states
(statistical operators) $\rho$ in
$C^2\otimes C^2$. We say that $\rho$ is
an MDS  state (one with maximally
disordered subsystems or rather
subsystem states) if $\rho_1=(1/2)I_1$
and $\rho_2=(1/2)I_2$. R. and M.
Horodecki have shown \cite{Hor} that
for every MDS state there exist unitary
subsystem operators $U_1$ and $U_2$
such that $$ \Big(U_1\otimes
U_2\Big)\rho \Big(U_1^{\dagger}\otimes
U_2^{\dagger} \Big)=(1/4)\Big(I\otimes
I+\sum_{i=1}^3 t_i \sigma_i\otimes
\sigma_i\Big)\equiv T, \eqno{(5)}$$
where $\sigma_i$, $i=1,2,3$, are the
well known Pauli matrices $\sigma_x,
\sigma_y$ and $\sigma_z$; and it is
seen from their place in the expression
if they are meant for the first or for
the second spin-one-half particle.

Further, they have shown that the
operator $T$ is a statistical operator
(a quantum state) if and only if the
vector $\vec t$ from {\bf R}$_3$ the
components of which appear in (5) is
not outside the {\it tetrahedron}
determined as the set of all mixtures
of the four pure {\it Bell states}:
$$\ket{\psi^1_2}\equiv
(1/2)^{1/2}\Big(\ket{+}\ket{+}\enskip
{\buildrel - \over +}\enskip \ket{-}
\ket{-}\Big), \quad
\ket{\psi^3_0}\equiv
(1/2)^{1/2}\Big(\ket{+} \ket{-}\enskip
{\buildrel + \over -}\enskip \ket{-}
\ket{+}\Big), \eqno{(6)}$$
 where $\ket{\buildrel + \over -}$ are the
spin-up and the spin-down state vectors
respectively.

It is straightforward to see that the
three nonsinglet Bell states
$\ket{\psi_s}$, $s=1,2,3$, when written
in the form (5), are given by $t_s=-1$,
and the other two components of $\vec
t$ equal to $+1$. The singlet state
$\ket{\psi_0}$ is in the form (5)
determined by all three components of
$\vec t$ being equal to $-1$.

It is also easy to see that for all
mixtures one has
$$ -1\leq t_i\leq
+1\qquad i=1,2,3.$$
This is a
necessary, but not a sufficient
condition for $T$ being a state. In
other words, the tetrahedron is
embedded in a cube, in which there are
also nonphysical $\vec t$. In view of
the LHS of (5), we call $T$ that belong
to the tetrahedron: {\it generating MDS
states}.

What we want to find out is: {\it Which
of the MDS  states have nontrivial
twins?} For those that do have, we want
to find {\it the set of all nontrivial
pairs of twins}.

The importance of the question, as it
was stated, lies in the fact that
nontrivial twins and only they make
possible distant orthogonal state
decomposition (measurement of $A_2$
without interaction, on account of
entanglement) by a measurement (of
$A_1$) on the nearby subsystem.

It is sufficient to find the generating
MDS  states $T$ with nontrivial twins,
because the validity of $$ A_1T=A_2T $$
obviously implies
$$
\Big(U_1A_1U_1^{\dagger}\Big)
\Big(U_1U_2TU_1^{\dagger}
U_2^{\dagger}\Big)=\Big(U_2A_2U_2^{\dagger}
\Big)\Big(U_1U_2TU_1^{\dagger}
U_2^{\dagger}\Big), $$
 i. e., if the
generating MDS  states have nontrivial
twins, then also the generated MDS
states do have nontrivial twins, and
they are immediately obtained.

As far as the pure generating MDS
states (the Bell states) are concerned,
the first-particle reduced statistical
operator $\rho_1$ is equal to
$(1/2)I_1$, {\it all} nontrivial
Hermitian operators $A_1$ commute with
it, hence, they are twins. To evaluate
the corresponding twin $A_2$, one has
to read off the antilinear correlation
operator $U_a$ from (6) having in mind
(2), and then utilize (4). For the best
known Bell state, the singlet state
$\ket{\psi_0}$, e. g., $U_a$ takes
$\ket{+}$ into $\ket{-}$, and $\ket{-}$
into $(-\ket{+})$ (cf (6)). If $$
A_1=\alpha_{++}\ket{+}\bra{+}\enskip
+\enskip
\alpha_{--}\ket{-}\bra{-}\enskip
+\enskip \alpha_{+-}
\ket{+}\bra{-}\enskip +\enskip
(\alpha_{+-})^*\ket{-} \bra{+},$$
$$\alpha_{++},\alpha_{--}\in \mbox{\bf
R},\quad \alpha_{+-}\in \mbox{\bf C},$$
 then the twin
$A_2$ has the form: $$
A_2=\alpha_{--}\ket{+}\bra{+}\enskip
+\enskip
\alpha_{++}\ket{-}\bra{-}\enskip
-\enskip \alpha_{+-}
\ket{+}\bra{-}\enskip -\enskip
(\alpha_{+-})^*\ket{-} \bra{+}.$$

Now we turn to the {\it mixtures of
Bell states} in our search for
nontrivial twins.\\

\section{Mixtures of
Bell states}

 Viewing statistical operators as super
vectors, and utilizing (redundantly,
but for the sake of better overview)
the ket notation for super state
vectors (i. e., Hilbert-Schmidt
operators as normalized super vectors),
one can rewrite the generating vectors
$T$ given by (5) as a biorthogonal
expansion with positive expansion
coefficients:

$$ \ket{T\|T\|^{-1}}_{12}=
(1+\sum_{i=1}^3t_i^2)^{-1/2}
\Big(\ket{(1/2)^{1/2}I}_1\otimes
\ket{(1/2)^{1/2}I}_2+$$
$$\sum_{i=1}^3|t_i|
\ket{(1/2)^{1/2}\sigma_i}_1 \otimes
\ket{sg(t_i)(1/2)^{1/2}\sigma_i}_2\Big)
\eqno{(7)}$$ ("$sg$" denotes the sign),
i. e., as a (super state vector) {\it
Schmidt canonical expansion}.

One can read off (7) the following
canonical entities of the super state
vector $ \ket{T\|T\|^{-1}}_{12}$:

The first-subsystem reduced statistical
super operator $\hat \rho_1$ has the
characteristic super state vectors
$\{\ket{(1/2)^{1/2}I}_1,\enskip
\ket{(1/2)^{1/2}\sigma_i}_1:\enskip
i=1,2,3\}$; the second subsystem
reduced statistical super operator
$\hat \rho_2$ has the characteristic
state vectors
$\{\ket{(1/2)^{1/2}I}_2,\enskip
\ket{sg(t_i)(1/2)^{1/2}
\sigma_i}_2:\enskip i=1,2,3\}$; and the
common spectrum of $\hat \rho_1$ and
$\hat \rho_2$ is $\{R_0\equiv
(1+\sum_{i=1}^3 t_i^2)^{-1},\enskip
R_i\equiv R_0t_i^2:\enskip i=1,2,3\}$.
Finally, the antiunitary correlation
super operator $\hat U_a$ maps the
enumerated characteristic state vectors
of $\hat \rho_1$ into the
correspondingly ordered ones of $\hat
\rho_2$.

\section{Nontrivial MDS  twins}

 Every super
operator $\hat A_1$ that commutes with
$\hat \rho_1$, i. e., for which every
characteristic subspace of the latter
is invariant (and no other super
operator ), has a twin super operator
$\hat A_2$. But we are interested only
in those pairs $(\hat A_1,\hat A_2)$,
{\it both} super operators, which are,
what may be called, {\it
multiplicative} ones, i. e., which have
the form $$ \hat
A_1\rho_{12}=A_1\rho_{12},\quad \hat
A_2\rho_{12}=A_2\rho_{12}, $$
 where $A_p$,
$p=1,2$, are ordinary (subsystem)
operators. It is easy to see that a
multiplicative super operator is {\it
Hermitian} (in the Hilbert-Schmidt
space of supervectors) if so is the
ordinary operator (in the usual sense)
that determines it.

The basic result of this study is given
in the following two theorems:\\

\noindent {\it Theorem 1.} Mixed
generating MDS states have nontrivial
twins if and only if they are mixtures
of two Bell states (binary mixtures).\\

\noindent {\it Theorem 2.} A) Let us
take a binary mixture of two Bell
states both distinct from the singlet
one, and let $T_i\equiv
\ket{\psi_i}\bra{\psi_i}$ (cf (6)) be
the nonsinglet Bell state that does not
participate in the mixture. Then the
nontrivial twins are:
 $$
A_1\equiv \alpha I_1 +\beta
\sigma_i^{(1)},\qquad A_2\equiv \alpha
I_2 +\beta \sigma_i^{(2)},\qquad \alpha
,\beta \in \mbox{\bf R},\enskip \beta
\not= 0, \eqno{(8)}$$
 where the suffix
on $\sigma_i$ refers to the
corresponding tensor factor space.

 B) In case of a binary mixture
of the singlet state with another Bell
state, say $T_i\equiv
\ket{\psi_i}\bra{\psi_i}$ (cf (6)), the
twins are:
 $$
A_1\equiv \alpha I_1 +\beta
\sigma_i^{(1)},\qquad A_2\equiv \alpha
I_2-\beta \sigma_i^{(2)},\qquad \alpha
,\beta \in \mbox{\bf R},\enskip \beta
\not= 0. \eqno{(9)}$$

Proof of the two theorems and of some
subsidiary results is given in the
Appendix.

All the binary generating MDS mixtures
$T^{(2)}$ appear to be nonseparable.
But this actually is not so.

It is shown elsewhere \cite{biort} that
a necessary and sufficient condition
for a {\it separable} composite-system
state to have {\it nontrivial twins} is
that the terms can be grouped into {\it
biorthogonal} groups of terms (as many
groups as there are detectable
characteristic values of the twin
Hermitian operator).

According to Theorem 1, all binary
generating MDS states $T^{(2)}$ do have
nontrivial twins. On the other hand, it
is easily seen that {\it only two} of
these states satisfy the mentioned
condition for separable states. As one
can easily ascertain making use of (6),
they are: $$
(1/2)\Big((\ket{+}\bra{+}\otimes
\ket{+}\bra{+})\enskip +\enskip
(\ket{-}\bra{-}\otimes
\ket{-}\bra{-})\Big)=$$

$$ (1/2)\Big(\ket{\psi_1}\bra{\psi_1}
+\ket{\psi_2}\bra{\psi_2}\Big),
\eqno{(10a)}$$ and $$
(1/2)\Big((\ket{+}\bra{+}\otimes
\ket{-}\bra{-})\enskip +\enskip
(\ket{-}\bra{-}\otimes \ket{+}\bra{+})
\Big)=$$ $$
(1/2)\Big(\ket{\psi_0}\bra{\psi_0} +
\ket{\psi_3}\bra{\psi_3}\Big).
\eqno{(10b)}$$

The proof of Theorem 2 (given in the
Appendix) is only of {\it
methodological significance}: it
illustrates a method how to evaluate
nontrivial twins. In our case of binary
mixtures $T^{(2)}$, another method
gives a simpler evaluation.\\

\noindent {\it Theorem 3.} A pair of
opposite-subsystem observables
$(A_1,A_2)$ are twins for a
composite-system mixture  $$ \rho
=\sum_nw_n\ket{\Psi_n}\bra{ \Psi_n}$$
{\it if and only if} they are
simultaneously twins {\it for each} of
the pure term states.\\

(This is one of the results of a
previous article \cite{HD00}, section
3., C2.) Proceeding as outlined in the
Introduction, it is straightforward to
evaluate the twins in the operator
basis consisting of the four
supervectors
$\ket{\stackrel{+}{-}}\bra{\stackrel{+}{-}}$.
But for comparison with the results (8)
and (9) obtained by the Schmidt
canonical expansion method, to which
this article is devoted, we do this in
a little bit more difficult way using
the form (7) for the Bell states (see
their description beneath (7)).

We can read off the antiunitary
correlation superoperator $\hat U_a$
from the mentioned form (7) of the Bell
state. As it was stated before, every
first-subsystem observable $A_1\equiv
\alpha I_1
+\sum_{i=1}^3\beta_i\sigma_i^{(1)},\enskip
(\alpha ,\beta_i\in \mbox{\bf
R},\enskip i=1,2,3)$, is a twin. The
corresponding second-subsystem twins
for the Bell states are: $$ T_1:\quad
A_2\equiv \alpha I_2
-\beta_1\sigma_1^{(2)}+\beta_2
\sigma_2^{(2)}
+\beta_3\sigma_3^{(2)};$$ $$ T_2:\quad
A_2\equiv \alpha I_2
+\beta_1\sigma_1^{(2)}-\beta_2
\sigma_2^{(2)} +
\beta_3\sigma_3^{(2)};$$ $$ T_3:\quad
A_2\equiv \alpha I_2+\beta_1
\sigma_1^{(2)}+ \beta_2\sigma_2^{(2)}-
\beta_3 \sigma_3^{(2)};$$ $$ T_0:\quad
A_2\equiv \alpha I_2-\beta_1
\sigma_1^{(2)}- \beta_2\sigma_2^{(2)}-
\beta_3 \sigma_3^{(2)}.$$

Now, in view of the position of the
minus sign in $A_2$, evidently,
utilizing $m\not= i\not= j\not=
m\enskip i,j,m\in \{1,2,3\}$, and
$0<w<1$, the simultaneous twins are: $$
wT_j+(1-w)T_m:\quad A_2\equiv \alpha
+\beta_i\sigma_i;$$ $$
wT_0+(1-w)T_i:\quad A_2\equiv \alpha
-\beta_i\sigma_i;$$  and $A_2$ is, of
course, the twin of $A_1\equiv \alpha
+\beta_i\sigma_i$.

In this way proof of (8) and (9) is
obtained.\\

\section{Appendix}

Since we are going to prove the
theorems making use of (7), first we
must be able to recognize the binary
mixtures $T^{(2)}$ on the Horodecki
tetrahedron.\\

\noindent {\it Proposition A.1.} One
has a binary mixture $T^{(2)}$ if and
only if precisely one of the three
$|t_i|$ values in (7) equals $1$.

A) If $t_i=+1,\enskip |t_{i+1}|,
|t_{i+2}|<1$ (where the three values
$\{1,2,3\}$ of $i$ are meant
cyclically), then the mixture is of two
Bell states both distinct from the
singlet state. If $T_i$ is the
nonsinglet Bell state that does not
participate in the mixture, one has
$t_{i+2}= -t_{i+1}$. Finally, the
binary mixture $T^{(2)}$ in question is
$$
T^{(2)}=\Big[(1-t_{i+1})/2\Big]T_{i+1}
+\Big[(1-t_{i+2})/2\Big]T_{i+2}.\eqno{(A1)}$$

B) If $t_i=-1,\enskip |t_{i+1}|,
|t_{i+2}|<1$ (in the cyclic sense),
then one deals with a mixture of two
states: the singlet state and another
Bell state $T_i$. One has $t_{i+1}
=t_{i+2}$, and the binary mixture
$T^{(2)}$ in question is $$
T^{(2)}=\Big[(1+t_{i+1})/2\Big]T_i+
\Big[
(1-t_{i+1})/2\Big]T_0.\eqno{(A2)}$$
Both in the cases (A) and (B),
$t_{i+1}$ can be any number in the
interval $-1\leq t_{i+1}\leq +1$;
equivalently, one can have any point on
the corresponding border of the
Horodecki tetrahedron (the vertices
excluded).\\

For proof a few subsidiary results are
required.\\

\noindent {\it Lemma A.1.} If among the
four numbers $\{1,|t_i|:i=1,2,3\}$
appearing in the form (7) of the
generating MDS state $T$ there is one
distinct from the rest, then $T$ has no
nontrivial twins.\\

\noindent {\it Proof.} As clearly
follows from the above stated spectrum
of $\hat \rho_1$, the mentioned "one
number distinct from the rest"
corresponds to a nondegenerate
characteristic value. Assuming that
$A_1$ is a twin, it is a multiplicative
superoperator reducing in each
characteristic subspace of $\hat
\rho_1$. (This is equivalent to
commutation with $\hat \rho_1$.)

    a) Let us take the case when
$|t_i|<1,\quad i=1,2,3$. Then the first
characteristic value of $\hat \rho_1$
is nondegenerate, and the corresponding
characteristic super state vector has
to be invariant (up to a constant): $$
A_1(1/2)^{1/2}I_1=\alpha
(1/2)^{1/2}I_1,$$
i. e., $A_1=\alpha$,
and the twin is trivial.

b) Let $|t_i|$ for some value of $i$ be
distinct from the other three numbers.
Then the corresponding characteristic
super state vector
$\sigma_i(1/2)^{1/2}$ must be invariant
(up to a constant):
$$
A_1\sigma_i(1/2)^{1/2}=\alpha
\sigma_i(1/2)^{1/2}, $$
 which, upon
multiplication with $\sigma_i$ from the
right, implies $A_1=\alpha$
again.\hfill $\Box$ \\

\noindent {\it Corollary A.1.} If a
generating MDS state $T$ has nontrivial
twins, then for at least one value of
$i$: $|t_i|=1$.\\

\noindent {\it Proof} is obvious from
Lemma A.1.\hfill $\Box$ \\

\noindent {\it Lemma A.2.} Expressing a
generating MDS  state $T$ written in
the form (7) in terms of the
statistical weights with respect to the
Bell states $\{T_k\equiv
\ket{\psi_k}\bra{\psi_k}:k=0,1,\dots
,3\}$ (cf (6)), one has: $$
T=\sum_{k=0}^3w_kT_k= (1/4)
\Big[I\otimes
I+(-w_1+w_2+w_3-w_0)\sigma_1\otimes
\sigma_1+(w_1-w_2+w_3-w_0)\sigma_2\otimes
\sigma_2+$$
$$(w_1+w_2-w_3-w_0)\sigma_3\otimes
\sigma_3\Big],\eqno{(A3)}$$
 where
$$ \forall k:\quad w_k\in [0,1],\quad
k=0,1,2,3;\quad \sum_{k=0}^3w_k=1.$$

\noindent {\it Proof} is
straightforward substituting the Bell
states in (7) (cf (6) and beneath
it).\hfill $\Box$ \\

\noindent {\it Lemma A.3.} If  one has
$|t_i|=1,\enskip i=1,2,3$, for a
generating MDS state $T$ in the form
(7), then it is a Bell state.\\

\noindent {\it Proof.} Each $t_i$ has
two sign possibilities; altogether
there are $2^3=8$ possibilities. A
straightforward analysis of each of
these, taking into account Lemma A.2
and $\sum_{k=0}^3w_k=1$, shows that $4$
possibilities do not give states. These
are: $\{sg(t_i)=+:i=1,2,3\},\enskip
\{+--\},\enskip \{-+-\}$, and
$\{--+\}$. The remaining four sign
possibilities give the four Bell
states:
$$ \{-++\}:\enskip T_1;\quad
\{+-+\}:\enskip T_2;\quad
\{++-\}:\enskip T_3;\quad
\{---\}:\enskip T_0.$$
 \hfill  $\Box$

\noindent {\it Proof of claim (A) in
Proposition A.1.}  Since it is clear
from (A3) that the $t_i$ as functions
of $w_k$ are symmetric (in the sense of
the cycle $\{1,2,3\}$), it is
sufficient to take $i=1$. Then
$$
-w_1+w_2+w_3-w_0=1,\quad \mbox{and}
\quad \sum_{k=0}^3w_k=1. $$
 This gives
$w_2+w_3=1,\enskip w_1=w_0=0$, and
$t_2= w_3-w_2=-t_3$. Hence, $w_2=
(1-t_2)/2$ and $w_3=(1+t_2)/2$ as
claimed. Since $0< w_1,w_0<1$, the
claimed intervals for $t_2$ and $t_3$
follow.\hfill $\Box$ \\

\noindent {\it Proof of claim (B) of
the Proposition.} It runs in full
analogy with the proof for case
(A).\hfill $\Box$ \\

\noindent {\it Proof of the main claim
of the Proposition.} It is easy to see
that the proofs of claims (A) and (B)
of the Proposition go through also for
$|t_{i+1}|$ or $|t_{i+2}|$ equalling
one. Hence, one cannot have $|t_i|=1$
for precisely two values of $i$. If it
is so for one value, then either it is
so for all three values - and one has a
pure Bell state, or it is so for
precisely one value of $i$, then we
have a binary mixture.\hfill $\Box$ \\

\noindent {\it Proof of Theorem 2.} We
now assume that for one value of $i$,
$|t_i|=1$, and that the other two
components of $\vec t$ in (7) are by
modulus less than one. Then it is
sufficient and necessary for an
observable $A_1$ that defines a
superoperator $\hat A_1$ by
multiplication (we write this as $\hat
A_1\equiv (A_1\bullet )$) to have a
superoperator twin $A_2$ (that is not
necessarily multiplicative as $\hat
A_1$) that it reduces in the
two-dimensional supervector subspace
spanned by $I_1$ and $\sigma_i^{(1)}$.
If we write $A_1=\alpha I_1
+\sum_{j=1}^3\beta_j\sigma_j^{(1)}$
($\alpha ,\beta_j\in${\bf R}), and
multiply with this from the left
$\sigma_i^{(1)}$, it turns out that the
condition amounts to $\beta_j=0,\enskip
j\not= i$. The symmetrical argument
gives the symmetrical result. Thus the
multiplicative superoperators defined
by $A_1$ and, separately, by $A_2$ do
have superoperator twins if and only if
they are of the form
$$A_1=\alpha
I_1+\beta \sigma_i^{(1)},\qquad
A_2=\gamma I_2 +\delta
\sigma_i^{(2)},\eqno{(A4)}$$
 where $\alpha
,\beta ,\gamma ,\delta \in \mbox{\bf
R}$.

 The mentioned operators are
twins of each other if and only if $$
(A_2\bullet )=\hat U_a(A_1\bullet )
\hat U_a^{-1}.\eqno{(A5)}$$

Now we find out the necessary and
sufficient conditions when (A5) is
valid for the operators given by (A4).
Since both sides of (A5) are linear
operators, we apply them to the basis
of supervectors
$\{I_2,\sigma_i^{(2)}:i=1,2,3\}$: $$
(A_2\bullet )I_2=\gamma I_2+\delta
\sigma_i^{(2)};$$
$$ \Big(\hat
U_a(A_1\bullet )\hat
U_a^{-1}\Big)I_2=\hat U_a(\alpha I_1+
\beta \sigma_i^{(1)})=\alpha I_2+
sg(t_i)\beta \sigma_i^{(2)}.$$

Thus, we obtain the condition
$$ \gamma
=\alpha ,\quad \delta= sg(t_i)\beta .$$

Utilizing the well known relation $$
\sigma_i\sigma_j=\delta_{ij}I+
\sum_{m=1}^3i\epsilon_{ijm}\sigma_m,$$
 we, further, have
$$ (A_2\bullet )\sigma_j^{(2)}=(\gamma
I_2+ \delta \sigma_i^{(2)})
\sigma_j^{(2)}=\gamma \sigma_j^{(2)}+
\delta (\delta_{ij}I_2+\sum_mi
\epsilon_{ijm}\sigma_m^{(2)});$$ $$
\Big(\hat U_a(A_1\bullet )\hat
U_a^{-1}\Big)\sigma_j^{(2)}=sg(t_j)\hat
U_a(\alpha I_1+\beta \sigma_i^{(1)})
\sigma_j^{(1)}=$$ $$ sg(t_j)\hat
U_a\Big( \alpha \sigma_j^{(1)}+\beta
(\delta_{ij}I_1+\sum_mi\epsilon_{ijm}
\sigma_m^{(1)})\Big)=$$
$$sg(t_j)\Big(\alpha sg(t_j)
\sigma_j^{(2)}+\beta (\delta_{ij}I_2-
\sum_mi\epsilon_{ijm}sg(t_m)
\sigma_m^{(2)})\Big).$$
 For $i=j$ we
obtain the condition $\gamma =\alpha$,
and $\delta =sg(t_i)\beta$, and, for
$j\not= i$, in addition: $\delta
=-sg(t_j) sg(t_m)\beta$. Since $i\not=
m\not= j$, we know from the Proposition
that, irrespective of $sg(t_i)$, one
has $-sg(t_j)sg(t_m)=sg(t_i)$. Hence,
we actually obtain the condition
expressed by (8) and (9).\hfill $\Box$

The claim in Theorem 1 that binary
mixtures $T^{(2)}$ do have nontrivial
twins is an immediate consequence of
Theorem 2.\\

\noindent {\bf Acknowledgements}\\

\noindent The author wants to thank
Prof. Anton Zeilinger for his
invitation as well as the "Erwin
Schr\"{o}dinger" Institute in Vienna,
where part of this work was done, for
the financial support and hospitality
extended to him.

\end{document}